\documentclass[12pt]{article}
\usepackage{a4}\usepackage{amsfonts}\textwidth15cm\textheight23.7cm
\usepackage{graphicx}
\newcommand{\Ref}[1]{(\ref{#1})}
\newcommand{\be}{\begin{equation}}\newcommand{\ee}{\end{equation}}
\newcommand{\bea}{\begin{eqnarray}}\newcommand{\eea}{\end{eqnarray}}
\newcommand{\beq}{\begin{eqnarray}}\newcommand{\eeq}{\end{eqnarray}}
\newcommand{\beao}{\begin{eqnarray*}}\newcommand{\eeao}{\end{eqnarray*}}
\newcommand{\tra}{\top}
\newcommand{\E}{{\cal E}}
\newcommand{\nn}{\nonumber}\newcommand{\pa}{\partial}
\newcommand{\al}{\alpha}\newcommand{\ep}{\epsilon}

\newcommand{\om}{\omega}\newcommand{\Om}{\Omega}

\newcommand{ \text}[1]{#1 }\newcommand{\tr}{{\rm tr}}
\newcommand{\BMK}{\cite{BORDAG2006D} }    
\newcommand{\plasmons}{\cite{Bordag:2005by} }   
\newcommand{\BHR}{\cite{Hennig:1990ap,Bordag:1992cm} }   
\newcommand{\BRW}{\cite{Bordag:1985zk} }
\begin{document}
\title{On the interaction of a charge with  a thin plasma sheet}
\author{
{\sc M. Bordag}\thanks{e-mail: Michael.Bordag@itp.uni-leipzig.de} \\
\small Halmstad University\\ \small Box 823, 30118 Halmstad, Sweden\footnote{Permanent address:  University of Leipzig, Institute for Theoretical Physics,
  Vor dem Hospitaltore 1, 04103 Leipzig, Germany}}
\date{}
\maketitle
\begin{abstract}
The interaction of the electromagnetic field with a two dimensional plasma sheet intended to describe the pi-electrons of a carbon nano-tube or a $C_{60}$ molecule is investigated. By integrating out first the displacement field of the plasma or first the electromagnetic field different representations for quantities like the Casimir energy are derived which are shown to be consistent with one another. Starting from the covariant gauge for the electromagnetic field it is shown that the matching conditions to which the presence of the plasma sheet can be reduced are different from the commonly used ones. The difference in the treatments does not show up in the Casimir force between two parallel sheets, but it is present in the Casimir-Polder force between a charge or a neutral atom and a sheet. At once, since the plasma sheet is a regularization of the conductor boundary conditions, this sheds light on the difference in physics found earlier in the realization of conductor boundary conditions as 'thin' or 'thick' boundary conditions in Phys.Rev.D70(2004)085010.
\end{abstract} 
\section{Introduction}

The boundary conditions the electromagnetic field has to obey on the surface of a conductor,
\be
\label{bc}\E_{||}=B_\perp=0,
\ee
belong to the standard topics in classical electrodynamics. However, in \cite{Bordag:2004dn} it was shown that the conditions \Ref{bc} do not fix the electromagnetic field completely, rather they allow for a freedom in the normal component of the electric field, $\E_\perp$, on the surface of the conductor. In the standard treatment, which is mostly done in Coulomb gauge, this freedom is fixed by applying Gauss's law, $\rm div \E=0$. In \cite{Bordag:2004dn} it was shown that   a consistent scheme can be constructed in which $\E_\perp$ may well stay without any condition on the surface. The difference between both treatments is that in \cite{Bordag:2004dn} the surface was assumed to be mathematically thin, i.e., purely two-dimensional, whereas in the standard treatment a conducting bulk behind the surface is assumed, from which the condition on the normal component might be deduced. For that reason, in \cite{Bordag:2004dn} the standard realization was called 'thick' conductor  and the new realization in which no additional condition on $\E_\perp$ are assumed  was called 'thin' conductor. The question to which extend the 'thin' conductor is physical, could not be answered in a satisfactory way in \cite{Bordag:2004dn} due to the lack of a physical model for an infinitely thin conducting sheet. However,         it is desirable to have such a model since in \cite{Bordag:2004dn} it was shown that 'thick' and 'thin' conductors result in different physics. For instance, for 'thin' conductors there is an interaction of classical charges across the surface which is absent for 'thick' conductor. Furthermore, the Casimir-Polder force between an atom and a surface comes out by 13 \% smaller for 'thin' as compared to 'thick'  conductor. It is interesting to note that the Casimir force between two flat surfaces is the same for both. It should be noticed that the difference between both approaches is not in the different gauges chosen but in the boundary conditions. In \cite{Bordag:2004dn} it was shown that in the approach with 'thin' boundary conditions and covariant gauge the standard result can be reproduced if imposing an additional boundary condition on the normal component of the electric field which does not follow from \Ref{bc}.

In the present paper we answer the question about a physical realization of the 'thin' conductor. As such we consider a two dimensional plasma sheet and show that it delivers in the  limit of its plasma frequency to infinity just the 'thin' conductor and, for instance, the reduced Casimir-Polder force. 
The two-dimensional plasma sheet plays a twofold role. On the one hand side it is a regularization of a ideally conducting surface (its plasma frequency $\Om$ plays the role of the regularization parameter) which is recovered in the limit $\Om\to\infty$. On the other hand side it is a model for the $\pi$-electrons of a graphene sheet, a $C_{60}$ molecule or a carbon nano-tube. The interaction of an atom with such a object is not only a measurable quantity but it is at present of obvious practical interest. 

In the next section we start from the general formulation of a system consisting of the electromagnetic field and an electric fluid (plasma) confined to some two-dimensional surface $S$. The fluid  is described by a displacement vector $\vec{\xi}$ and we use the formulation in terms of a functional integral representation which is, as usual, not the only possible, but the most convenient formulation. We work in covariant gauge which in general terms looks more natural even in connection with the non-relativistic dynamics of the fluid. 
In order to make the treatment more convincing we consider two different ways by first integrating out either the displacement field 
$\vec{\xi}$ or the electromagnetic field. In the first case we obtain the photon propagator in the presence of the plasma sheet and in the second one we get the propagator of the displacement field. In both cases we consider the excitation spectrum, i.e., the plasmons. To simplify calculations we first consider a scalar model before entering the  discussion of the photon polarizations. As applications we (re-)calculate the Casimir force between two plasma sheets and the Casimir-Polder force between an atom and one sheet. In the appendix we consider the excitations for a spherical plasma shell. \\
Throughout the paper we use units with $\hbar=c=1$.
\section{The plasma shell model and its interaction with the electromagnetic field}
The plasma shell model describes an electrically charged fluid confined to a two-dimensional surface $S$ by a displacement vector $\vec{\xi}(x)$ ($x\in S$) which is tangential to $S$. A immobile, overall electrically neutralizing background is assumed. This model was probably first considered in \cite{Fetter73} to describe a layered electron gas and, later, in \cite{Barton:1991pb} to describe the plasmon oscillations in $C_{60}$. In fact, this simple model describes quite well a number of properties of the $\pi$-electrons in $C_{60}$, carbon nano-tubes and objects alike \cite{Maksimenko2002}. With respect to the Casimir effect it was considered in a series of papers, \cite{BIV} and successors. In \plasmons the interaction of two such sheets with emphasis on the role of the surface plasmons and in \BMK the interaction of one such sheet with a dielectric half space and with a neutral atom were calculated. 
As for the past two papers it should be mentioned that there the standard realization of boundary conditions was taken as granted and that both papers started from the boundary conditions on the TE and TM modes and standard polarizations. As it will be shown below the results of \plasmons remain valid whereas that of \BMK need to be reconsidered. At least the result obtained there for the interaction of an atom with a graphene, i.e., with a plasma sheet,  is different from what we will find below in section 5. 

The action of the considered model reads
\bea
\label{S}
S&=&-\frac14   \int d^4x\ \left( F_{\mu\nu}^2 +\frac{1}{2\al}\left(\pa_\mu A_\mu\right)^2\right)
+ \frac{m}{2} \int_S d^3z \ \dot{\vec{\xi}~}\! (z)^2
\nn \\ && 
+ \int d^4x\  A_\mu(x)\left(j_\mu(x)+J_\mu(x)\right),
\eea
where $F_{\mu\nu}=\pa_\mu A_\nu-\pa_\nu A_\mu$ is the usual field strength tensor (we use the notation $\pa_\mu=\pa/\pa x_\mu$) and $\vec{\xi}$ is the displacement field of the fluid having a mass density $m$. We use $x,y,\dots$ to denote coordinates in the bulk and $z,z',\dots$ for coordinates on the surface $S$ and we adopt the usual summation convention that for repeated indices the sum over their range is assumed. So the first term in S, \Ref{S}, is the usual action of the electromagnetic field in covariant gauge with gauge fixing parameter $\al$, the second is the kinetic energy of the fluid (the dot denotes the time derivative and we assume  non-relativistic dynamics) and the third describes the interaction of the electromagnetic field with the fluid current $j_\mu(x)$ and some external current $J_\mu(x)$. The fluid current is
\bea
\label{j1}
j_0(x)&=&e\int_S dz\ \delta(x-f(z)) \ \nabla\vec{\xi}(z),
\nn \\\vec{
j}(x)&=&e\int_S dz\ \delta(x-f(z)) \ \dot{\vec{\xi}~}\!(z),
\eea
where $e$ is the charge density. The surface is described by $S=\left\{x\mid x=f(z)\right\}$ and $z$ is a parameterization of $S$. In the simplest case of a plane perpendicular to the $x_3$-axis at $x_3=0$, a parameterization is  $z=\{z_0,z_1,z_2\}$  and $f(z)=\{z_0,z_1,z_2,0\}$ defines the surface $S$. In that case the current is
\be
\label{j2}
j_\mu(x)=e\delta(x_3)\left( \begin{array}{c}\nabla\vec{\xi}(x)\\\dot{\vec{\xi}~}\!(x)\end{array}\right)_{\!\mu}
\ee
and the displacement vector is $\vec{\xi}(x)=\left(\begin{array}{c}\xi_1(x_\al) \\ \xi_2(x_\al) \end{array}\right)$ ($\al=0,1,2$). Current conservation holds, $\pa_\mu j_\mu=0$.

For the following it is convenient to represent the interaction in terms of an integral kernel, $H_{\mu ;i,b}(x,z)$,
\be
\label{H1}
\int d^4x\ A_\mu(x) j_\mu(x)\equiv 
\int d^4x\ \int_S d^3z\ A_\mu(x)
H_{\mu ;i,b}(x,z) \xi_{i,b}(z).
\ee
Obviously, this is possible for any surface $S$ and it allows, in this way, to consider the most general case. However, here we restrict ourselves to the special case of flat surfaces, located at $x_3=a_b$ ($b$ numbers the surfaces if more than one),
\be
\label{H2}
H_{\mu ;i,b}(x,z)=e\delta(x_\al-z_\al)\delta(x_3-a_b)
\left(\delta_{\mu 0}\pa_i-\delta_{\mu i}\pa_0 \right)
\ee
($\mu=0,\dots,3$, $\al=0,1,2$, $i=1,2$, $b=1,2$).
In the case of two surfaces the parameterization must include the number of the surface too. For a single surface the index $b$ can be simply dropped. 

Now let us consider the generating functional of the Greens functions,
\be
\label{Z}
Z(J)=\int DA \ D\xi \ \ e^{i S}
\ee
with the action $S$ rewritten in the form
\bea
\label{S1}
S&=&\frac12\int d^4x\ A_\mu (x)K_{\mu\nu} A_\nu(x)
+ \frac12 \int_S d^3z\  \xi_{i,b}(z)K^0 \ \xi_{i,b}(z)
\nn \\ &&
+\int d^4x \int_S d^3z\ A_\mu(x)  H_{\mu ;i,b}(x,z) \xi_{i,b}(z)
+
\int d^4x\ A_\mu(x) J_\mu(x),
\eea
where
\be
\label{K}
K_{\mu\nu}=g_{\mu\nu}\pa^2-\left(1-\frac1\al\right)\pa_\mu\pa_\nu
\ee
is the kernel of the free action of the electromagnetic field and
\be
\label{K0}
K^0= -m\pa_0^2 
\ee
is  for the displacement field. 

The generating functional is in the known way by means of 
\be
\label{F}
F=\frac{-i}{T}\ln Z(0)
\ee
connected with the free energy $F$ ($T$ is the total time and it drops out in a few steps). The distance dependent part of the free energy is then the vacuum or Casimir energy of the considered system.

At this place a remark on the role of the displacement field $\xi$ must be added. In letting its mass $m$ go to zero the term with the kinetic energy disappears and the action $S$, Eq.\Ref{S1}, becomes linear in $\xi$. Then the functional integral over $\xi$ can be carried out and it delivers a functional delta function, $\prod\limits_{{z,z'\in S\atop i=1,2;b=1,2}}\delta(\int dx\ A_\mu(x)H_{\mu;i,b}(x,z))$. As it will be apparent below, this is nothing else that the  treatment of conductor boundary conditions by functional delta functions introduced in \BRW. In this sense the kinetic energy of the displacement field provides a regularization of these delta functions and of the conductor boundary conditions. 

In the functional integral \Ref{Z}, the action $S$, Eq.\Ref{S1}, is quadratic in the fields, hence the integral is Gaussian and can be carried out. For the sake of clarity we adopt a symbolic notation dropping all  indices and arguments. The the action is
\be
\label{S1s}
S=\frac12 \ A\ K\ A+\frac12\ \xi \ K^0 \ \xi +A\ H\ \xi+AJ.
\ee
Now we {\it first integrate out the displacement field $\xi$}. We write  $S$ as complete square in $\xi$,
\be
\label{S2}
S=\frac12 \ A\ ^S\!K\ A+\frac12\ (\xi+\tilde{\xi}) \ K^0 \ (\xi+\tilde{\xi})+AJ,
\ee
with
\be
\label{}\tilde{\xi}=(K^0)^{-1}H^\top A
\ee
and
\be
\label{sK}^S\!K=K-H(K^0)^{-1}H^\top .
\ee
Here, $H^\top$ is the transposed in the sense that $AH\xi=\xi H^\top A$ must hold. The inverse $(K^0)^{-1}$ of $K^0$ like similar inverse below must be taken in the space in which the operator acts, the surface $S$ in this case.

In a second step we complete the square for the field $A$ in order to integrate out the electromagnetic field and represent the action in the form
\be
\label{S2a}
S=\frac12 \ (A+\tilde{A})\ ^S\!K\ (A+\tilde{A}) 
+\frac12\ (\xi+\tilde{\xi}) \ K^0 \ (\xi+\tilde{\xi})
-\frac12\ J (^S\!K)^{-1}J
\ee
with $\tilde{A}=(^S\!K)^{-1} J$. Now the Gaussian integrations can be carried out, first that over $\xi$, subsequently that over $A$ and we arrive at
\be
\label{Z1}
Z=\left(\det {^S\!K}\right)^{-1/2} \ \left(\det K^0\right)^{-1/2} \ e^{-\frac{i}{2} J (^S\!K)^{-1}J}.
\ee
In these formulas, ${^S\!K}$ is the kernel of the action of the electromagnetic field after integrating out the displacement field. i.e., taking into account the interaction with the plasma sheet. In general, we denote by an index $S$ in front  a kernel that  includes the interaction and by the index $0$ we mark a kernel belonging to the displacement field. 

From the action \Ref{S2}, the equations of motion for the field $A$ follow,
\be
\label{eomA}  {^S\!K} A=J,
\ee
where the interaction with the fluid is in the second term in the r.h.s. of ${^S\!K}$, Eq.\Ref{sK}. For vanishing source $J=0$, these are the  equations determining the modes of the electromagnetic field in the presence of the fluid. Below, in an example, we will see that ${^S\!K}$ provides the usual free space Maxwell equations supplemented by the matching conditions on the surface $S$. 

In this formulation, the Casimir energy is given according to \Ref{F} by 
\be
\label{F1}
F=\frac{i}{2T}\ \tr \ln {^S\!K}+\frac{i}{2T}\ \tr \ln K^0,
\ee
where the traces are taken in the corresponding spaces. For a setup  with two sheets the distance dependent part is contained in the first term, i.e. in $\tr \ln {^S\!K}$, whereas the second term, $\tr \ln K^0$, adds only a (infinite) constant. 

Now we consider the case of {\it first integrating out the electromagnetic field $A$}. For this, we rewrite the action \Ref{S1s} in the form 
\be
\label{S3}
S=\frac12 \ (A+\hat{A})\ K\ (A+\hat{A}) 
+\frac12\ \xi \ K^0 \ \xi
-\frac12\ \hat{A}\ K\ \hat{A},
\ee
where now
\be
\label{}\hat{A}=K^{-1}\left(H\xi+J\right).
\ee
Here $K^{-1}$ is the inversion of the kernel of the free action in the empty space, i.e., the free space photon propagator $D=K^{-1}$, in detailed writing
\be
\label{Dmn}D_{\mu\nu}(x-y)=\int\frac{d^4k}{(2\pi)^4}\ \frac{e^{-k_\mu(x_\mu-y_\mu)}}{k_0^2-\vec{k}^2+i0}
\left(-g_{\mu\nu}+\left(1-\al\right)\frac{k_\mu k_\nu}{k^2}\right)
\ee
($+i0$ for the causal propagator). After that we complete the square for $\xi$. By means of 
\be
\label{}\frac12\ \hat{A}\ K\ \hat{A}=
\frac12 \xi H^\top K^{-1}H\xi+\xi H^\top K^{-1}J+\frac12 JK^{-1}J
\ee
we get
\be
\label{S4}
S=\frac12 \ (A+\hat{A})\ K\ (A+\hat{A}) 
+\frac12\ \xi \ ^S\!K^0 \ \xi
-\xi H^\tra K^{-1}J
-\frac12\ J\ K^{-1}\ J
\ee
with 
\be
\label{sK0}^S\!K^0=K^0-H^\top K^{-1}H.
\ee
Obviously, $^S\!K^{0}$ is the kernel of the action of the displacement field after integrating out the electromagnetic field and  the equations of motion for the fluid read
\be
\label{eomxi}^S\!K^0 \ \xi =0
\ee
(below we will investigate this equation in detail). The first term in $^S\!K^0$, Eq.\Ref{sK0}, results from the kinetic energy and the second from the electromagnetic interaction of the fluid with itself, for instance, due to the Coulomb interaction. 

Completing now in Eq.\Ref{S4} the quadratic form for $\xi$ we arrive at 
\be
\label{S5}
S=\frac12 \ (A+\hat{A})\ K\ (A+\hat{A}) 
+\frac12\ (\xi+\hat{\xi})  \ ^S\!K^0 \ (\xi+\hat{\xi})
-\frac12\ J\ ^S\!D\ J
\ee
with $\hat{\xi}=(^S\!K^0)^{-1}H^\top K^{-1}J$ and
\be
\label{sD}
^S\!D=D-DH(^S\!K^0)^{-1}H^\top D,
\ee
which is the photon propagator in the presence of the fluid. Now the Gaussian integrations in \Ref{Z} can be carried out using the action \Ref{S5}, first that over the A-field, and then that over the $\xi$-field,
\be
\label{Z2}
Z=\left(\det K\right)^{-1/2} \ \left(\det {^S\!K}^0\right)^{-1/2} \ e^{-\frac{i}{2} J ^S\!DJ}.
\ee
The corresponding Casimir energy is
\be
\label{F2}
F=\frac{i}{2T}\ \tr \ln K+\frac{i}{2T}\ \tr \ln {{^S\!K}}^0.
\ee
Thanks to the derivation, this expression must be a representation of the same Casimir energy as in \Ref{F1}. The distance dependent part in \Ref{F2} is in the second term, whereas the first is distance independent. 

As seen, \Ref{F1} and \Ref{F2} are two different representations of the same quantity. This follows from the derivation. For instance, if taking a finite dimensional example where the kernels are represented by finite dimensional matrices acting in different spaces, this equivalence can be shown quite easily. In field theory where these kernels are infinite dimensional matrices, it is not so easy mainly due to the inherent divergences. However, there is no doubt that correct definitions can be given and that the equivalence can be shown. We are not going to do that and restrict ourselves to some examples.

First of all, let us compare the source terms in \Ref{Z1} and \Ref{Z2}. Since they must be equal,
\be
\label{sKsD}   {^S\!K}^{-1}={^S\!D},
\ee
${^S\!D}$ is the inversion of the kernel of the electromagnetic field after integrating out the displacement field. In this form  ${^S\!D}$ describes the modification of the photon propagator which comes in from the surface $S$. In fact, an expression of this type was derived in \BRW in the simpler case of conductor boundary conditions and in \BHR for delta-function potentials (later it was re-obtained by other authors and, for  instance in \cite{WIRZBA2006} it was discussed in connection with the Krein formula). 

Now we consider ${^S\!K}$ in the l.h.s. of Eq.\Ref{sKsD} in the form given by Eq.\Ref{sK}. Up to the Lorentz indices it consists of the wave operator and a second term which due to the delta functions in $j_\mu$, Eq.\Ref{j1} or \Ref{j2}, has support only on the surface $S$. It can be considered as a  potential with localized support and in the simplest case we get a wave equation with a delta function potential like the one considered in \BHR. Since such potentials are equivalent to known matching conditions the electromagnetic field has to obey on the surface $S$, the kernel $^S\!K$ is equivalent to the free wave equation with the corresponding boundary conditions. In this way Eq.\Ref{sKsD} can be verified.

Now we turn to the discussions of the expressions \Ref{F1} and \Ref{F2} for the Casimir energy and consider, say for two plasma sheets, the distance dependent part. In \Ref{F1} it results from $^S\!K$ and in \Ref{F2} from $^S\!K^0$. We note that these are different operators, acting in different spaces, $^S\!K$ in the bulk and $^S\!K^0$ on the surface $S$. Also their physical interpretation is different. $^S\!K$describes the photon fluctuation in the presence of the surface $S$ and $^S\!K^0$ describes the fluctuations of the displacement field $\xi$ including its electromagnetic self-interaction. The contribution from both to the Casimir energy must be the same. Moreover, this contribution enters only one time, either through $^S\!K$ in representation \Ref{F1} or through $^S\!K^0$ in representation \Ref{F2}, but not twice. This is quite counterintuitive since one is tempted to argue that both kinds of  fluctuations should  contribute to the vacuum energy. But since that is not the case one can conclude that in a scheme of canonical quantization it would not be possible to introduce independent creation and annihilation operators to both kinds of excitations. 

\section{A scalar field interacting with plane plasma sheets}
In order to have an explicit realization of the general formulas developed in the preceding section we consider the simplest non-trivial example, a scalar field interacting with one or two parallel plasma sheets. 
We start from the action \Ref{S1} and replace   the electromagnetic field $A$ by a real scalar field $\phi$,
\bea
\label{S1sc}
S&=&\frac12 \ \int d^4x\ \phi(x)\ K\ \phi(x)
+\frac12\int_S d^3z\ \xi_{i,b}(z) \ K^0 \ \xi_{i,b}(z) 
\nn \\ &&
+\int d^4x\ \phi(x) \left(j_0(x)+J(x)\right).
\eea
The kernel of the free action of $\phi$ is now simply the wave operator,
\be
\label{wo}K=-\pa^2=-\pa_0^2+\Delta \ ,
\ee
$K^0=-m\pa_0^2$ is the analogous quantity for the displacement field and the current is given by
\be
\label{jsc}j(x)=e\sum_{b=1,2}\delta(x_3-a_b)\nabla_i\xi_{i,b}(x).
\ee
In the case of two plasma sheets these are located at $x_3=a_b$ ($b=1,2$). In this model, $e$ can be viewed simply as a coupling constant without having the meaning of an electric charge and we define
\be
\label{Om}\Om=\frac{e^2}{m}
\ee
as plasma frequency. Strictly speaking, $e$ and $m$ are the corresponding densities per unit area of the sheets, however, since  only the ratio \Ref{Om} enters the final formulas we do not need to introduce separate notations for the densities.  The displacement of the fluid is described by a two dimensional vector $\xi_{i,b}(x_\al)$ in the plane of the sheet. Its components are labeled by $i=1,2$, $x_\al$ with $\al=0,1,2$ are the coordinates in the sheet and $b=1,2$ labels the sheets. The interaction between the field $\phi$ and the displacement field $\xi$ can be rewritten in the form of Eq.\Ref{H2},
\be
\label{}\int d^4x\ \phi(x) j(x)=\int d^4x\ \int d^3z\ \phi(x) H_{i,b}(x,z)\xi_{i,b}(z)
\ee
with
\be
\label{}
H_{i,b}(x,z)=e \delta(x_\al-z_\al)\delta(x_3-a_b)\nabla_i \ .
\ee
The transposition of $H$,
\be
\label{}H_{i,b}^\top(x,z)=-e \delta(x_\al-z_\al)\delta(x_3-a_b)\nabla_i
\ee
follows from
\be
\label{}
\int d^4x\ \int d^3z\ \phi(x) H_{i,b}(x,z)\xi_{i,b}(z)=
\int d^4x\ \int d^3z\ \xi_{i,b}(z) H_{i,b}^\top(x,z)\phi(x)
\ee
with integrating by parts.

For the simple geometry of two parallel sheets all these quantities can be simplified by Fourier transform in the $x_\al$-directions ($\al=0,1,2$). We define for the fields
\be
\label{Four}\phi(x)=\int\frac{d^3k_\al}{(2\pi)^3}\ e^{ik_\al x_\al} \ \phi_k(x_3)
\ee
and for the kernels
\be
\label{} K(x,y)=\int\frac{d^3k_\al}{(2\pi)^3}\ e^{ik_\al (x_\al-y_\al)} \ K_k(x_3,y_3)
\ee
and similar for $\xi$ and $K^0$. The transformed quantities are marked by an index $k$. Since the fields are real, their Fourier transforms obey $\phi_k(x_3)^*=\phi_{-k}(x_3)$.
The wave operator is connected with the corresponding kernel by $K(x,y)=\delta(x-y)K(x)$ and after Fourier transform it becomes
\be
\label{}K_k(x_3)=\Gamma^2+\pa_{x_3}^2
\ee
with
\be
\label{Ga}\Gamma\equiv \sqrt{k_\al^2+i 0}=\sqrt{k_0^2-k_1^2-k_2^2+i 0} \ ,
\ee
where we introduced the infinitesimal imaginary part  for later use. The corresponding quantity for the displacement field is simply
\be
\label{}K^0_k=mk_0^2
\ee
and for the interaction kernel we note
\be
\label{Hsc}{H_k}_{i,b}=iek_i\delta(x_3-a_b).
\ee
Another simplification occurs if introducing polarizations for the displacement field.
Obviously there are two polarizations,
\be
\label{xiEM} 
\xi_i=
\left(\begin{array}{c}k_1\\k_2\end{array}\right)_{\!i} \frac{1}{k_{||}}\ \xi^{\rm TE}_k+
\left(\begin{array}{c}-k_2\\k_2\end{array}\right)_{\!i} \frac{1}{k_{||}}\ \xi^{\rm TM}_k \ .
\ee
We use the notations TE and TM because the in the electromagnetic case these modes couple just to the corresponding polarizations.
From the explicit form \Ref{Hsc} it is clear that because of
\be\label{Hsc1}
{H_k}_{i,b}\xi_{i,b}=iek_{||}\delta(x_3-a_b)\xi^{\rm TE}_b
\ee
only the TE-mode of the displacement field couples to the scalar field. The effective coupling constant is $ek_{||}$.

Now we integrate out the displacement field $\xi$. The first quantity we have to consider is the inversion of $K^0$ entering Eq.\Ref{sK}. Using the Fourier transform this is simply
\be
\label{} (K^0)^{-1}=\int\frac{d^3k_\al}{(2\pi)^3}\ e^{ik_\al (x_\al-y_\al)} \ (K^0_k)^{-1}
\ee
with
\be
\label{}(K^0_k)^{-1}=\frac{1}{mk_0^2+i0}.
\ee
Here and in the following we always chose the causal propagator. 
Applying $H$ from the right and from the left and inserting into \Ref{sK} and using
 ${^S\!K_k}(x_3,y_3)\delta(x_3-y_3)={^S\!K_k}(x_3)$ we get
\be
\label{kd}
^S\!K_k(x_3)=\Gamma^2+\pa_{x_3}^2
-\frac{\Om k_{||}^2}{k_0^2+i0}\sum_{b=1,2}\delta(x_3-a_b),
\ee
where $k_{||}=\sqrt{k_1^2+k_2^2}$ is the momentum parallel to the sheets. 

Now we consider the equations of motion for the field $\phi$ which are analogous to Eq.\Ref{eomA}. Without source these read
\be
\label{eomphi}^S\!K_k(x_3)\ \phi_k(x_3)=0
\ee
and, rewritten,
\be
\label{}
\left(-\pa_{x_3}^2
+\frac{\Om k_{||}^2}{k_0^2+i0}\sum_{b=1,2}\delta(x_3-a_b)\right)\phi_k(x_3)
=\Gamma^2 \phi_k(x_3).
\ee
Obviously, this is nothing else than a Schr\"odinger equation with delta function potential, a favorite textbook example. In the given case the potential is repulsive  ($\Om>0$ follows from \Ref{Om} independently on whether $e$ has an interpretation as electric charge or not)  and the field $\phi$ has a continuous spectrum. 

It is well known that the delta-function potential can be reformulated in terms of matching conditions on the surface. The field must be continuous and its derivative has a jump,
\be
\label{mcsc}\mbox{discont } \phi'=\frac{\Om k_{||}^2}{k_0^2} \ \phi \ .
\ee
In quantum field theory such potentials were considered several times. We refer here to \BHR,  where delta potentials on two parallel planes were considered for the scalar field and also for a spinor field. There the propagators were written down and, using the energy-momentum tensor,  the Casimir force between two planes was calculated. The latter is equivalent to calculate the Casimir force using Eq.\Ref{F1} so that we do not need to repeat that calculation here.

Now we consider the second approach where the scalar field $\phi$ is integrated out first. The Fourier transform of the free space propagator follows by performing the integration over $k_3$ in \Ref{Dmn} (dropping the bracket containing the vector structure),
\be
\label{}D_k(x_3-y_3)=\frac{e^{i\Gamma\mid x_3-y_3\mid}}{2i\Gamma}
\ee
and from \Ref{sK0} we get
\be
\label{}
({^S\!K^0_k})_{i,b;j,b'}=mk_0^2\delta_{ij}\delta_{b,b'}-e^2\frac{k_ik_j}{2i\Gamma} \ e^{i\Gamma \mid a_b-a_{b'}\mid},
\ee
where we added the necessary indices. This kernel is defined on the surface $S$, hence after Fourier transform it is purely algebraic, i.e., it does not contain any derivative or dependence on the coordinate $x_3$. The corresponding equation of motion for the displacement field is given by Eq.\Ref{eomxi}. It describes the excitations of the plasma including its selfinteraction caused by the scalar field. To consider these excitations it is sufficient to consider one sheet. In that case and after division by $m$  the equations of motion read
\be
\label{}  \left(k_0^2\delta_{ij} -\Om\frac{k_ik_j}{2i\Gamma}\right)\xi_j=0 \ .
\ee
These are diagonalized by the polarizations \Ref{xiEM} and the eigenvalue equations become 
\bea
\label{eom1}k_0^2  -\Om\frac{k_{||}^2}{2i\Gamma} &=&0, \quad \mbox{(TE)}
\nn \\
k_0^2&=&0. \quad \mbox{(TM)}
\eea
The meaning of the second equation is simply that this mode does not couple to the scalar field and it does not have excitations. The equation for the first mode has no solution for $\Om>0$ (taking into account that $\Gamma$ has a positive imaginary part).

Now we consider the construction of the   propagator $^S\!D$ for the $\phi$-field which is given by Eq.\Ref{sD} in general terms. In fact this repeats the corresponding calculation in \BHR. Therefore we restrict ourselves here to the simplest case of having only one plane. In that case,  after Fourier transform it becomes   
\be
\label{eom2}^S\!D_k(x_3,y_3)=
\frac{e^{i\Gamma\mid x_3-y_3\mid}}{2i\Gamma}
-\frac{e^{i\Gamma\mid x_3\mid}}{2i\Gamma} \ 
iek_i \ (^S\!K^0_k)^{-1}_{ij} \ iek_j  \ 
\frac{e^{i\Gamma\mid y_3\mid}}{2i\Gamma}
\ee
with the inversion
\be
\label{}(^S\!K^0_k)^{-1}_{ij}=
\frac{1}{k_{||}^2}
\left(\begin{array}{c}k_1\\k_2\end{array}\right)_{\!i}
\left(\begin{array}{c}k_1\\k_2\end{array}\right)_{\!j} \
\frac{1}{mk_0^2-e^2\frac{k_{||}^2}{2i\Gamma}}
+
\frac{1}{k_{||}^2}
\left(\begin{array}{c}-k_2\\k_1\end{array}\right)_{\!i}
\left(\begin{array}{c}-k_2\\k_1\end{array}\right)_{\!j} \
\frac{1}{mk_0^2},
\ee
which is done using the polarizations from \Ref{xiEM}. The second polarization does not contribute and we get finally
\be
\label{sDksc}^S\!D_k(x_3,y_3)=
\frac{e^{i\Gamma\mid x_3-y_3\mid}}{2i\Gamma}
-\frac{e^{i\Gamma(\mid x_3\mid+\mid y_3\mid )}}{2i\Gamma} \ r(k)
\ ,
\ee
where
\be\label{refl}  r(k)=
\frac{1}{1-\frac{2i\Gamma}{\Om}\frac{k_0^2}{k_{||}^2}} 
\ee
has the meaning of being the reflection coefficient in the corresponding scattering problem. 
This propagator \Ref{sDksc} obeys the matching conditions \Ref{mcsc} and up to notations it coincides with the scalar propagator derived in \BHR.

In this way we demonstrated how the general formulas of section 2 can be realized in the simplest case of a scalar field interacting with plane plasma sheets. Especially the equivalence of different representations was demonstrated. 
\section{The electromagnetic field interacting with plasma sheets}
In this section we consider the electromagnetic field interacting with one or two flat plasma sheets. On the one hand side this is a straight forward generalization of the scalar case considered in detail in the preceding section, but on the other hand side from the electromagnetic case the most interesting conclusions follow. 

We start with introducing polarizations for the electromagnetic field. We do not chose the standard polarizations but those first introduced in \cite{Bordag:1984ht} (and later in 
\cite{Bordag:1998sw} for a sphere and a cylinder). After Fourier transform \Ref{Four} the polarizations vectors read
\be\label{Ems} 
E_\mu^0=\frac{1}{\Gamma}\left(\begin{array}{c}
                    k_0\\[6pt]k_1\\[6pt]k_2\\[6pt]0\end{array}\right), \
E_\mu^3=\left(\begin{array}{c}0\\[6pt]0\\[6pt]0\\[6pt]1\end{array}\right), 
\
E_\mu^1=\frac{1}{k_{||}}\left(\begin{array}{c}0\\k_2\\-k_1\\0\end{array}\right), 
\
E_\mu^2=\frac{1}{\Gamma k_{||}}\left(\begin{array}{c}
                        k_{||}^2\\k_0k_1\\k_0k_2\\0\end{array}\right).
\ee
These  form a basis,
\be
\label{Ediag}g_{\mu\nu}=E_\mu^s \ g_{st}\ E_\nu^t
\ee
with $g_{st}=\mbox{diag}(1,-1,-,1-1)$. Note that these vectors do not contain $k_3$ and that in the case of conductor boundary conditions only the polarizations with $s=1,2$ are affected by the boundary.  In the approach of the 'thin' conductor,   those with $s=0,3$ were kept free of conditions. We will see that this takes place also in the case of the plasma sheet. 

With \Ref{Ems} and the Fourier transform \Ref{Four}, the expansion of the electromagnetic potentials reads
\be
\label{mexp} A_\mu(x)=\int\frac{d^3k_\al}{(2\pi)^3}\ e^{ik_\al x_\al} \
\sum_{s=0}^3 E_\mu^s \ A_k^s(x_3),
\ee
where we showed the sum over $s$ explicitly. We note that without the plasma sheet the equations of motion for the amplitudes are $\left(\Gamma^2+\pa_{x_3}^2\right)A_k^s(x_3)=0$ with $\Gamma$ defined in  \Ref{Ga} and that the free part of the action becomes diagonalized when  inserting \Ref{mexp} into \Ref{S1}. We start from inserting \Ref{mexp} into the interaction with the plasma sheet, i.e., into \Ref{H1}. After Fourier transform we note (for one plasma sheet)
\be
\label{}(H_k)_{\mu,i}=ie\delta(x_3)\left(\delta_{\mu0}k_i-\delta_{\mu i}k_0\right)
\ee
and the together with the polarization vectors we get
\be
\label{HE}
E_\mu^s (H_k)_{\mu,i}=ie\delta(x_3)
\left\{\begin{array}{cl}
0,&(s=0,3) \\ [5pt]
\frac{k_0}{k_{||}} \left(\begin{array}{c}-k_2\\k_1\end{array}\right)_{\!i} ,&(s=1)
 \\ [18pt]
\frac{-\Gamma}{k_{||}} \left(\begin{array}{c}k_1\\k_2\end{array}\right)_{\!i} .&(s=2)
\end{array}\right.
\ee
In this way, only the polarizations \Ref{Ems} with $s=1,2$ couple to the displacement field of the plasma sheet. This is the most important observation since it allows to diagonalize the action completely and to establish the correspondence with the 'thin' conductor approach in \cite{Bordag:2004dn}. Using the mode expansion \Ref{xiEM} of the displacement field we get for the interaction \Ref{H1}
\be
\label{H3}  \int d^4x\ A_\mu(x) j_\mu(x) =
\int\frac{d^3k_\al}{(2\pi)^3}\  ie \ \left(
k_0 \ \xi^{\rm TE}_k \ {A_k^1}(0)^* -\Gamma \ \xi^{\rm TM}_k \ {A_k^2}(0)^*
\right).
\ee
Here ${A_k^s}(0)^*$ ($s=1,2$) are the complex conjugated amplitudes taken at the position of the sheet, i.e., at $x_3=0$. In this way the action is diagonalized and the interaction of the electromagnetic field with the plasma sheet is reduced to two scalar problems. More exactly, the complete action becomes a sum of thee parts. The first contains the polarizations with $s=0,3$. This part does not depend on the plasma sheet. It contains the gauge fixing parameter $\al$. The next two parts, with $s=1$ and $s=2$, are just the two scalar problems where the plasma sheet enters and these are independent on the gauge fixing parameter $\al$. For them we can use the results of the preceding section with the substitution of the effective coupling constants introduced in Eq.\Ref{Hsc1} according to
\bea
\label{effk}    ek_{||}&\to&ek_0 ,   \qquad \mbox{(TE)} \nn \\
                ek_{||}&\to&e\Gamma .\qquad ~\mbox{(TM)}
\eea
These substitutions follow from comparing Eq. \Ref{Hsc} with \Ref{H3}. 
In order to use them for the photon propagator we represent it in the form
\be
\label{sD1}    
^S\!D_{\mu\nu}(x,y)=D_{\mu\nu}^{(s=0,3)}(x,y)+\sum_{s=1,2}D_{\mu\nu}^{(s)}(x,y).
\ee
Here the first part carries the gauge dependence and it is independent on the plasma sheet. The remaining two parts read in Fourier representation
\be
\label{sD2}
D_{\mu\nu}^{(s)}(x,y) = \int\frac{d^3k_\al}{(2\pi)^3}\  e^{ik_\al x_\al} \
E_\mu^s  E_\nu^s   \ ^S\!D^{(s)}_k(x_3,y_3)
\ee
($s=1,2$, no sum over $s$ here) with
\bea
\label{sD3}
^S\!D^{(s)}_k(x_3,y_3) &=&\frac{e^{i\Gamma\mid x_3-y_3\mid}}{2i\Gamma}
- \frac{e^{i\Gamma ( \mid x_3\mid +\mid y_3\mid)}}{2i\Gamma} \ 
 r_s(k)
\nn \\ &\equiv& D^{(s)}_k(x_3,y_3) -\overline{D}^{(s)}_k(x_3,y_3),
\eea
where in the last line the separation into free part and boundary dependent addendum was introduced which we will use in the next section. The reflection coefficients $r_s(k)$ contain  the different couplings \Ref{effk} relative to the scalar case  which can be accounted for by the formal substitutions of the plasma frequencies
\be
\label{Oms} \Om_s=  \left\{ 
\begin{array}{cl}
\Om\frac{k_0^2}{k_{||}^2} ,&(s=1)
 \\ [12pt]
\Om\frac{\Gamma^2}{k_{||}^2} .&(s=2)
\end{array}
\right.
\ee
With these relations, the reflection coefficients are
\bea
\label{r12}
r_{1}(k)&=&\frac{1}{1-\frac{2i\Gamma}{\Om}}  , \nn \\
r_{2}(k)&=&\frac{1}{1-\frac{2i k_0^2}{\Om\Gamma}} . 
\eea
From a comparison with \cite{BV} it follows that $r_1(k)$ is just the reflection coefficient for the TE-mode which is completely obvious and that $r_2(k)$ is the reflection coefficient for the TM-mode which becomes obvious after the remark the by virtue of the equations of motion $\Gamma$ is the same as a momentum perpendicular to the sheet if one introduces such. 

The difference between the treatments of the scalar problems corresponding to the TM-mode here and in the standard approach is the following.
The matching condition for the TM-mode in the approach taken here implies  that the mode function is continuous across the sheet and that its derivative has a jump given by Eq.\Ref{mcsc} with $\Om$ substituted by $\Om_s$ from \Ref{Oms} and in this way it corresponds to a delta function potential. In the standard approach one introduces another  mode function whose derivative is continuous and whose value across the sheet has a jump. That corresponds to a potential given by the derivative of a delta function. As for the resulting scalar problems, these are obviously the same as far as they depend only on the reflection coefficient $r_2(k)$, Eq.\Ref{r12}. From here as a consequence it follows that the Casimir force between two sheets is the same in both approaches. Remember that it can be calculated  by a generalized Lifshitz formula where only the reflection coefficients enter. This is also the reason why the results of \plasmons remain valid.

However, this does not imply that the Casimir force between a plasma sheet and a dielectric half space must be the same in both approaches since for the dielectric half space one would use the standard polarization and for the plasma sheet the new ones. 

In the standard approach it is known that for the TM mode there is a surface plasmon and that in the TE mode there is none. The same hold in the present approach too. For the TE mode the matching condition follows from \Ref{mcsc} with the upper line of \Ref{Oms} and it reads simply
\be
\label{mcTE}\mbox{discont } \phi'=\Om   \ \phi \ ,
\ee
where $\phi$ stands for $A_k^1(x_3)$ introduced in Eq.\Ref{mexp}.
This corresponds to a repulsive delta potential. In a similar way, for the TM mode we get with the lower line in \Ref{Oms},
\be
\label{mcTM}\mbox{discont } \phi'=\Om \frac{ \Gamma^2}{k_0^2} \ \phi \ ,
\ee
where $\phi$ stands for $A_k^2(x_3)$ introduced in Eq.\Ref{mexp}.
As can be seen, for $k_0^2<k_{||}^2$ the delta potential becomes attractive. In that case  the one-dimensional problem has one bound state and this bound state is a wave propagating along the surface in the three dimensional case.  These surface plasmons were considered   for instance in \cite{BV}. It should be mentioned that these are to a large extend analogs of the  surface plasmons known to travel on a flat surface of a dielectric body if for the permittivity the plasma model is used. The same statements can be obtained from the behavior of the reflection coefficients \Ref{r12} on the upper imaginary momentum axis. 

There is another way to consider these waves. Namely, let us consider the excitations of the plasma, i.e., the excitations of the displacement field. These are subject to the general equation \Ref{eomxi}. Knowing  that we have two scalar problems for the amplitudes $\xi^{\rm TE}_k$ and $\xi^{\rm TM}_k$ of the displacement field, which were introduced in Eq.\Ref{xiEM}, with plasma frequencies given by Eq.\Ref{Oms}, we get from the corresponding scalar equation, namely from the upper line in \Ref{eom1}, the equations
\bea
\label{eomTEM}
k_0^2  -\Om\frac{k_{0}^2}{2i\Gamma} &=&0, \quad \mbox{(TE)}
\nn \\
k_0^2  -\Om\frac{\Gamma}{2i} &=&0. \quad \mbox{(TM)}
\eea
For the TE-case we note that for $\Om>0$ because of the positive imaginary part of $\Gamma$ we do not have a solution besides the trivial one $k_0=0$. In opposite, for the TM-case we have a solution for $k_0^2<k_{||}^2$, 
\be
\label{disprel}k_0^2=\frac{\Om}{8}\sqrt{\Om^2+16k_{||}^2}-\frac{\Om^2}{8}.
\ee
We note that this spectrum is just the same as for the surface plasmon following from Eq.\Ref{mcTM}  mentioned above. There is reason to think that this coincidence is not accidental. Let is return to section 2. The equations of motion for the displacement field are given by Eq.\Ref{eomxi}. After diagonalization, the zeros of $^S\!K^0$ define the spectrum. On the other hand side, the spectrum of the electromagnetic field can be found from the poles of its propagator $^S\!D$ which in the presence of the plasma sheet is given by Eq.\Ref{sD}. There in the right hand side we have $^S\!K^0$ in the denominator so that it zeros define poles of $^S\!D$. In this way one can show in the general case that the surface plasmons and the excitations of the plasma (which are usually called plasmons)  have the same spectrum, or more exactly, that for each mode of the displacement excitations there is a corresponding mode in the electromagnetic spectrum. In the Appendix we illustrate this on the example of a spherical plasma sheet. 
%
\section{Interaction of a charge with a plasma sheet}
In this section we consider the interaction of a charge and of a neutral atom with a plane plasma sheet. For the neutral atom this is the same setup as for the Casimir-Polder force in \cite{Cp48}. We use also the same method, the quantum mechanical perturbation theory. Since the original paper, the Casimir-Polder force was considered repeatedly, see for instance \cite{Barton:1970hw} or \cite{Hinds}. 
In all cases, for the photon propagator the same setup was taken which in \cite{Bordag:2004dn} was called 'thick' boundary condition. In \cite{Bordag:2004dn}, the Casimir-Polder force was calculated for the 'thin' boundary conditions and found to be smaller. In this section we repeat that calculation for a plasma sheet in place of the conducting surface used in \cite{Bordag:2004dn}. It will be shown that we reproduce the result for 'thin' boundary conditions in the limit of infinite plasma frequency $\Om$. 

The calculation of the Casimir-Polder force is quite tedious and involves a number of subtle moments. All these are known in literature and we do not need to repeat them here. Instead, we follow very closely the derivation given in \cite{Bordag:2004dn} and focus on the modifications which come in from the plasma sheet in place of the conducting surface. In the initial setup of the problem we have three systems, the electromagnetic field, the displacement field and the charge resp. the atom whose interaction with the plasma sheet is the subject of our interest. We eliminate the displacement field by, say integrating it out in a functional integral setup like in section 2. In this way we assume that there is no other interaction between the charge and the plasma sheet  than through the electromagnetic field.  After that we we are left with the photon propagator in the presence of the sheet, i.e., obeying matching conditions as discussed in the preceding section. This propagator is given by Eq.\Ref{sD} in general terms and by Eqs.\Ref{sD1} to \Ref{sD3} for a flat sheet. With this photon propagator we repeat the calculation done in section 2 in \cite{Bordag:2004dn}. We note for instance that we use here the same polarizations $E_\mu^s$ and that the whole difference is in the presence of the reflection coefficients $r_{1,2}(k)$, Eq.\Ref{r12}, which are now present in the photon propagator. For the conducting surface, of course, $r_{1,2}=1$ holds which for the plasma sheet is recovered in the limit $\Om\to\infty$. 

The calculation starts from a quantum mechanical Hamilton operator, Eq.(26) in \cite{Bordag:2004dn},
\be\label{Hnr} H=\frac{\left(\vec{p}-e\vec{A}\right)^2}{2m}+e
A_0+V(x), \ee
for a particle (electron) with momentum $p$ in a potential (of the nucleus) $V(x)$  and interacting with the electromagnetic potential $A_\mu(x)$. We note that $e$ and $m$ in \Ref{Hnr} do not need to be the same as for the plasma sheet whose properties enter only through its plasma frequency $\Om$, \Ref{Om}. 

The first quantity to be calculated is the static interaction of the charge with the nucleus, or more exactly, the change of this interaction brought in by the plasma sheet. In \cite{Bordag:2004dn} this was $\Delta V$, Eq.(31), which could there be written down directly as the potential of the mirror charge of the nucleus. In general it is given as the zeroth component of the field generated by a static source. In terms of the propagator this is
\be
\label{DV}\Delta V(x)=e^2\int d^4y\ \overline{D}_{00}(x,y) \delta(\vec{y})\ ,
\ee
where we included only the boundary dependent part $ \overline{D}_{00}(x,y)$ of the propagator which in our case is given by Eq.\Ref{sD3}. In momentum representation, the integration over $y_0$ makes the momentum $k_0$ vanish and with formulas \Ref{sD1} to \Ref{sD3} we obtain
\be
\label{DV1}\Delta V(x)=e^2\int\frac{dk_{||}}{(2\pi)^2}\ e^{ik_{||}x_{||}} \ 
 \frac{k_{||}^2}{\Gamma^2} \ (-r_2(k)) \ \frac{e^{i\Gamma(2a-x_3)}}{2i\Gamma}.
\ee
Here we have taken into account that following \cite{Bordag:2004dn} the sheet is located at $x_3=a$ and the atom in the origin. A contribution to the $(0,0)$ component in the boundary dependent part  comes only from the polarization vector $E_\mu^2$ in \Ref{Ems} and it delivers   the factor $k_{||}^2/\Gamma^2$. Further we have to take into account that for $k_0=0$ we have $\Gamma=ik_{||}$ so that this factor simply becomes (-1). The reflection coefficient $r_2(k)$, given by Eq.\Ref{r12} simplifies too, for $k_0=0$ it becomes $r_2(k_0=0)=1$. That means we get the same result as for the ideal conductor. In fact, performing the integration in \Ref{DV1} we just get
\be\label{DV2} \Delta
V(x)=\frac{e^2}{4\pi\sqrt{x_1^2+x_2^2+(x_3-2a)^2}}\ ,
\ee
and further, as a first order perturbation of  the static  energy\footnote{Following the notations in \cite{Barton:1970hw} and \cite{Bordag:2004dn}, till the end of this section interaction energies are denoted by $\delta$.}, 
\be
\label{}\delta_{\rm es,~ nucl}=\langle n\mid \Delta V(x) \mid n\rangle=\frac{e^2}{4\pi}\left(\frac{1}{2a}+\frac{Q}{16a^3}+\dots\right)
\ee
follows, where $Q$ is the quadruple moment. 

The next step is to calculate the corrections following from $\vec{A}(x)^2$-term in  the Hamilton operator. Its boundary dependent term $\delta_1$ is given by Eq.(36) in \cite{Bordag:2004dn},
\be\label{d1a}\delta_1=\frac{e^2}{2m} \ \frac{1}{i}
<n|\overline{D}_{ii}(x,x)|n>.\ee
We insert the propagator into this formula and pay attention to the contributions from the two polarizations,

\be\label{d1c}
\delta_1=\frac{e^2}{2m}\frac{1}{i}\int\frac{d^3k_\al}{(2\pi)^3} \
\frac{-e^{2i\Gamma a}}{-2i\Gamma}
\left(r_1(k)+r_2(k)\frac{k_0^2}{\Gamma^2}\right) <n\mid e^{2i\Gamma
(x^3-a)}\mid n >. \ee
In this formula one needs to perform the Wick rotation, $k_0\to ik_4$, $\Gamma\to i\gamma\equiv i\sqrt{k_4^2+k_{||}^2}$ and after expanding the matrix elements we get in generalization of Eq. (38) in \cite{Bordag:2004dn}
\bea\label{d1d} \delta_1&=-&\frac{e^2}{2m}\int\frac{d^3_{\rm E}
k}{(2\pi)^3} \ \frac{e^{-2\gamma a}}{2\gamma}
\left(\tilde{r}_1+\tilde{r}_2 \frac{k_4^2}{\gamma^2}\right) 
 \eea
with
\be
\label{rtilde}\tilde{r}_1=\frac{1}{1+\frac{2\gamma}{\Om}}, \qquad \tilde{r}_2=\frac{1}{1+\frac{2k_4^2}{\Om\gamma}}.
\ee
Since the integrations in this formula cannot be performed explicitly we represent $\delta_1$ in the form
\be
\label{delta1}\delta_1=
-\frac{e^2}{4\pi} \ \frac{1 }{8\pi
ma^2}\left(f_{\rm TE}(\Om a)+\frac{1}{3}f_{\rm TM}(\Om a)\right)+O\left(\frac{1}{a^4}\right) \ ,
\ee
where the functions
\bea
\label{}
f_{\rm TE}(x)&=&\int_0^\infty dk \ \frac{k \ e^{-k}}{1+k/x}\ , \nn \\
f_{\rm TM}(x)&=&3x\int_0^\infty dk \   e^{-k} \ \left(1-\sqrt{\frac{x}{k}}\arctan\sqrt{\frac{k}{x}}\right)\ ,
\eea
describe the reduction of the corresponding contribution due to the plasma sheet relative to the ideal conductor case. We note
\be
\label{}f_{\rm TE}(\infty)=f_{\rm TM}(\infty)=1
\ee
and, for $x\to 0$,
\be
\label{}f_{\rm TE}(x)\sim x+\dots, \qquad f_{\rm TM}(x)\sim 3x+\dots \ .
\ee

Next we consider the second order perturbation which in \cite{Bordag:2004dn} is given by Eq.(41). Again, the change coming in from the plasma sheet is to insert the reflection coefficients into the contributions from the corresponding polarizations. Proceeding directly to the expression (43) in \cite{Bordag:2004dn} the second order perturbation reads now
\be\label{d2c}      \delta_2
=\sum_{n'}\sum_{s=1}^2  \ \frac1i
\int\frac{d^3k_\al}{(2\pi)^3} \frac{-e^{2i\Gamma a}}{-2i\Gamma} \ r_s(k) \ 
\frac{\mid<n\mid \tilde{ G}_s\mid
n'>\mid^2}{-k_0+E_n-E_{n'}(1-i\ep)} \ .\ee
For the matrix elements the expressions (43) and (44) in \cite{Bordag:2004dn} remain valid.

The calculation advances with considering separately the cases of a {\it single charge} and a {\it neutral atom} in front of the sheet. For a single charge one uses the so called 'no-recoil' approximation which implies an expansion for small energy differences $E_n-E_{n'}$ in the denominator,
\be\label{smalla}\frac{1}{-k_0+E_n-E_{n'}}=
-\frac{1}{k_0}-\frac{E_n-E_{n'}}{k_0^2}-\frac{(E_n-E_{n'})^2}{k_0^3}+\dots
\, . \ee
Using the expansion of the matrix elements given in \cite{Bordag:2004dn} and the sum rules listed in \cite{Bordag:2004dn} in the appendix  we come to
\bea
\label{}
\delta_2^{\rm no-recoil}&=&
 \frac1i
\int\frac{d^3k_\al}{(2\pi)^3} \frac{-e^{2i\Gamma a}}{-2i\Gamma}\ \frac{-1}{k_0} \ 
\left\{ e^2 r_2(k)\left(-1+\frac{k_0^2}{\Gamma^2}\right)
\right. \nn \\ &&\left.
~~~~~~~~~+\frac{e^2}{m^2}\left[ r_1(k)+r_2(k)\left(\frac{k_0^2}{\Gamma^2}+\frac{k_{||}^4}{k_0^2\Gamma^2}\right)\right]
        \frac12\langle  p_{||}^2\rangle 
\right. \nn \\ &&\left.
~~~~~~~~~+\frac{e^2}{m^2}  \frac{-k_{||}^2}{k_0^2} \   \langle  p_{3}^2\rangle 
\right\} \ .
\eea
Now the integration can be carried out as described by  Eqs.(49) in \cite{Bordag:2004dn} as pole contribution  delivering
\bea
\label{dnrf}
\delta_2^{\rm no-recoil}&=&
\frac12\int\frac{dk_{||}}{(2\pi)^2} \ \frac{e^{-2k_{||}a}}{2k_{||}} \ 
\Bigg\{ -e^2
\nn \\ &&
~~~~~~~~+\frac{e^2}{m^2} \left[\frac{1}{1+\frac{2k_{||}}{\Om}}-\left(ak_{||}+\frac32
    +\frac{2k_{||}}{\Om}\right)\right]\frac12\langle  p_{||}^2\rangle
\nn \\ &&
~~~~~~~~~+\frac{e^2}{m^2} \left(-ak_{||}-\frac12-\frac{k}{\Om}\right) 
        \langle  p_{3}^2\rangle      \Bigg\}
\nn
\\
&=&\frac{-e^2}{8\pi a}-\frac{e^2}{m^2}
\left[ h_{||}(\Om a) \frac12\langle  p_{||}^2\rangle +
        h_3(\Om a) \langle  p_{3}^2\rangle  \right]
\eea
with 
\bea
\label{}
h_{||}(x)&=& \int_0^\infty dk\ e^{-k} \ \left(\frac{-1}{1+k/x}+\frac{k}{2}+\frac32+\frac{k}{x}\right),
\nn \\
h_3(x)&=&1+\frac{1}{x} \ .
\eea
This is the interaction between a single charge and the plasma shell. The first term in \Ref{dnrf} is the electrostatic contribution which does not depend on the plasma frequency $\Om$ and it coincides with the ideal conductor limit. The second contribution depends on $\Om$. For $\Om\to\infty$ it turns into the ideal conductor case calculated in \cite{Bordag:2004dn} for the 'thin' boundary conditions. The weakening of this interaction energy relative to the conductor case is described by the functions $h_{||}(x)$ and $h_{3}(x)$ which both obey $h_{||}(\infty)=h_{3}(\infty)=1$. In the opposite limit, i.e., for small $\Om$, these functions diverge which means that this limit is in conflict with the approximations made in deriving \Ref{dnrf}.

We note that the interaction between a single charge and a plasma sheet was calculated in \cite{BVI}, section 4, within the standard approach using Coulomb gauge.The result is different from \Ref{dnrf} in the part proportional to $\langle  p_{||}^2\rangle$ (the contribution proportional to $\langle  p_{3}^2\rangle$ is the same). This result was to be expected since in \cite{BVI} the ideal conductor limit is what in \cite{Bordag:2004dn} is called the 'thick' conductor.  It must be mentioned that, as discussed in \cite{Bordag:2004dn}, the 'thin' conductor boundary conditions were derived for an infinitesimal thin conducting sheet. The question whether these apply to a conducting surface with a conducting bulk behind is still opened. It can be expected that in that case rather the 'thick' conductor boundary conditions apply. However, this question needs further investigation. In opposite, the interaction with the plasma sheet, as follows from the above calculations, is uniquely described by \Ref{dnrf}.

Now we turn to a neutral atom in front of the plasma sheet. The starting point is Eq.\Ref{d2c}. This integral is split into two parts according to Eq.(51) in \cite{Bordag:2004dn}. To the first part, which is the pole contribution in $k_0=0$, only the intermediate state with $n=n'$ contributes. Since the momentum operator $p$ in the matrix elements (see Eqs.(43) and (44) in \cite{Bordag:2004dn}) has no diagonal contribution we have a non-vanishing contribution from the second polarization only. But the reflection coefficient $r_2(k)$, \Ref{r12}, is for $k_0=0$ equal to unity. Hence this contribution is the same as for ideal conductor, ie.e, it is the same as calculated in \cite{Bordag:2004dn}, Eq.(55), and it cancels against   other static contributions. The second part in \Ref{d2c} results from the VP-integral (see Eq. (51) in \cite{Bordag:2004dn}) and after performing the Wick rotation it becomes
\be\label{d2nd1} \delta_2^{\rm VP}=
-\mbox{VP}\int\frac{d^3_{\rm E}k}{(2\pi)^3} \frac{e^{-2\gamma a}}{2\gamma}
\sum_{n'}\sum_{s=1,2}  \ \tilde{r}_s \ \frac{\mid <n\mid \hat{G}_s\mid n'
>\mid|^2}{-ik_4+E_n-E_{n'}} \ , \ee
where $\tilde{r}_s$ are given by Eq.\Ref{rtilde}.
In the next step we expand the denominator for small $k_4$ and use again the sum rules listed in the appendix in \cite{Bordag:2004dn}. After that $\delta_2^{\rm VP}$ can be represented in the form
\bea
\label{}\delta_2^{\rm VP}&=&
\int\frac{d^3_{\rm E}k}{(2\pi)^3} \frac{e^{-2\gamma a}}{2\gamma}
\Bigg\{   \frac{e^2}{m}\left(\tilde{r}_1(k)+\tilde{r}_2(k)\frac{k_4^2}{\gamma^2}\right)
\nn \\ && ~~~~~~~~~~
-\left(\tilde{r}_1(k) k_4^2+\tilde{r}_2(k) \frac{k_4^4+k_{||}^4}{\gamma^2} \right)
\frac{\al_1+\al_2}{4}
-\tilde{r}_2(k)k_{||}^2 \ \frac{\al_3}{2}
\Bigg\}\ ,
\eea
where $\al_i$ is the static polarizability of the atom in the $i$-th direction. Finally we use spherical coordinates in the $k$-integrations with $\ep=\cos\theta$,
\bea
\label{}\delta_2^{\rm VP}&=&\frac{1}{4\pi^2}\int_0^\infty dk \ k\ e^{-2ak}\int_0^1 d\ep\       \Bigg\{
\frac{e^2}{m}\left(\tilde{r}_1+\tilde{r}_2 \ep^2\right)
\nn \\ &&
-k^2  \left(\tilde{r}_1\ep^2+\tilde{r}_2\left(\ep^4+(1-\ep^2)^2\right)\right)
            \frac{\al_1+\al_2}{4}
            -\tilde{r}_2(1-\ep^2) \frac{\al_3}{2}   \Bigg\} \ .
\eea
Here the first contribution cancels against $\delta_1$, \Ref{delta1}, and the rest is what contributes to the Casimir-Polder force. It can be written in the form
\be
\label{CPgra}\delta_{\rm CP}=\frac{-1}{32\pi^2a^4} \left\{
\left(g_{\rm TE}(\Om a)+\frac{11}{5}g_{\rm TM}(\Om a)\right)\frac{\al_1+\al_2}{4}
+g_3(\Om a) \ \al_3\right\} \ ,
\ee
where the functions
\bea
\label{}
g_{\rm TE}(x)&=&\frac16\int_0^\infty dk\ \frac{k^3 \ e^{-k}}{1+k/x} \ ,  \nn \\
g_{\rm TM}(x)&=&\frac{5}{22}\int_0^\infty dk\ k^3 \ e^{-k}\int_0^1 d\ep \ \frac{\ep^4+(1-\ep^2)^2}{1+\ep^2 k/x} \ ,  \nn \\
&=&\frac{5}{22}\int_0^\infty dk\ k^3 \ e^{-k}
 \left( \frac{2x}{3k}-\frac{2x(x+k)}{k^2}
    +\frac{\sqrt{x}(2x^2+2xk+k^2)}{k^{5/2}}\arctan\sqrt{\frac{k}{x}} \right)  \nn \\
g_{3}(x)&=&\frac{1}{4}\int_0^\infty dk\ k^3 \ e^{-k}\int_0^1 d\ep \ \frac{1-\ep^2}{1+\ep^2 k/x} \nn \\
&=&\frac{1}{4}\int_0^\infty dk\ k^3 \ e^{-k}  \left(-\frac{x}{k}+\frac{\sqrt{x}(x+k)}{k^{3/2}}\arctan\sqrt{\frac{k}{x}} \right) 
\ ,\eea
describe the decrease of the corresponding contributions relative to the ideal conductor case considered in \cite{Bordag:2004dn}. All these functions are normalized to unity for $x\to\infty$. Also, they vanish for $\Om\to0$ so that within the approximations made the interaction with the plasma sheet disappears in this limit.  

The interaction of an atom with a plasma sheet was considered in \BMK, Eq.(25), following the standard treatment and, consequently, it comes out to be different from \Ref{CPgra}.

\section{Conclusions}
In the foregoing sections we considered a plasma sheet interacting with the electromagnetic field. We used a representation in terms of a functional integral. From that it was seen that the fluid acts as a regularization for the conductor boundary conditions which are obtained in the limit of $\Om\to\infty$, where $\Om$ is the plasma frequency of the fluid. The consideration of this plasma shell model was  motivated by the wish to have a physical model for the realization of the 'thin' boundary conditions introduced in \cite{Bordag:2004dn} for a infinitely thin conducting sheet. Indeed, the plasma shell model may serve as such since it is a meaningful model for the $\pi$-electrons of  a graphene sheet. For the electromagnetic field we used the same  polarizations $E_\mu^s$, \Ref{Ems}, as for the 'thin' boundary conditions in \cite{Bordag:2004dn}. From this and from the reflection coefficients \Ref{r12},  it is clear that in the   limit $\Om\to\infty$ we get just    the 'thin' boundary conditions. It should be underlined that this is insofar a nontrivial result as this derivation is straightforward and it does not leave room for any
 freedom or arbitrariness. 
 This follows from the procedure applied  consisting in 
 first integrating out the displacement field which is done without any assumption on the gauge fixing or the polarizations of the electromagnetic field. After that we have an action for the electromagnetic field with the kernel $^S\!K$, Eq.\Ref{sK} and the corresponding equations of motion, \Ref{eomA}. This kernel is uniquely defined and cannot be altered. 
The 'remaining task', then, is the solution of the equations \Ref{eomA}. 
 We use polarizations which diagonalize this kernel $^S\!K$ and come to two separated scalar problems. The one (TE-mode) is identical to the corresponding one in the standard approach, the other one (TM-mode) is different in that it has a different polarization vector (the corresponding reflection coefficients are the same). 

 We would like to note that the attempt to solve the mentioned equations using the standard polarizations would hit the problem that these do not diagonalize $^S\!K$ as can be seen easily. Also, we do not see how the standard transition to Coulomb gauge can be done here since the equation of motion for the component $A_0$ of the vector potential is not instantaneous (in opposite to the case when there is no displacement field). 
At the moment we do not have a good explanation what might be the problem within the standard approach. We restrict ourselves here to the repetition of  the remark done already in the Introduction that  in \cite{Bordag:2004dn} we have shown that   the standard approach, i.e., the 'thick' boundary conditions,  follow from the 'thin' boundary conditions by introducing by hand a restriction on the normal component of the electric field in addition to \Ref{bc}.

We would like to summarize the most sensible conclusions as follows:
\begin{enumerate}
\item For an ideal conducting body, i.e., for a surface with  boundary conditions $\E_{||}=B_\perp=0$ and a conducting bulk behind (this is the standard treatment, we call it 'thick boundary conditions'), our results do not apply. 
\item For a {\it infinitely thin} surface, the boundary conditions $\E_{||}=B_\perp=0$ can be realized in a different way which we called 'thin boundary conditions'. These describe different physics, for instance, a by $\sim$13\% reduced Casimir-Polder force. 
\item For a two dimensional plasma sheet we get a unique description which is different from the standard one and which delivers, for instance, a different Casimir-Polder force. In the limit of infinity plasma frequency when the sheet becomes a  infinitely thin conducting surface, we obtain the thin boundary conditions. 
\end{enumerate}

As for the physical consequences in relation to the standard approach we showed that the Casimir force between two plasma sheets is the same whereas the Casimir-Polder force between an atom (or a charge) and a plasma sheet is different in both approaches. We cannot by-pass the conclusion that our approach gives the correct result.

In the second section we considered the calculation of the functional integral (it is Gaussian) in two ways, by first integrating out either the displacement field or the electromagnetic field. Different representations follow whose comparison allows for some insights. For instance, for the Casimir energy we obtained two different representations. In \Ref{F1} the distance dependent part of the Casimir energy can be attributed to the vacuum fluctuations of the electromagnetic field whereas in \Ref{F2} to those of the displacement field. In fact, this is a statement on what are the independent degrees of freedom, namely either the first or the second ones but not both at the same time. 
Furthermore, since the same Casimir energy can be calculated either from the modes living in the bulk or from modes living on the surface one could think of some kind of holographic principle. 

Another observation concerns the excitations. It turns out that the surface plasmons of the electromagnetic field have the same spectrum as the excitations of the fluid (which are usually considered the plasmons). We confirm the known results that there is no plasmon for the TE mode and that there is one (for each plasma shell) for the TM mode which has the same dispersion relation as known from \cite{BV}. In the Appendix we considered a spherical plasma sheet confirmed the corresponding conclusions (there is no plasmon at all).

We have seen that there exist two realizations of the boundary conditions $\E_{||}=B_\perp=0$. For the 'thin boundary conditions', by means of the plasma shell, we have a regularization. The regularization parameter is the plasma frequency $\Om\to\infty$. Hence, it would be useful to have a similar regularization for the 'thick boundary conditions'. One could imagine a plasma shell of finite thickness (which is a well investigated topic), allowing for a displacement $\vec{\xi}$ (in the bulk) not only parallel but also normal to the surface. This displacement field could be integrated out delivering a photon propagator which could be investigated in two different limiting procedures. The one would be   to remove the regularization, i.e., to let $\Om\to\infty$, which should deliver the 'thick boundary conditions'. The other way should be first to shrink the thickness of the plasma sheet to zero and after that to let $\Om\to\infty$.  This should  deliver the 'thin boundary conditions'.

\section*{Acknowledgements}
The author is  indebted to G. Barton for discussions on the topic which contributed   to a better understanding.
However, we still disagree in the treatment of the problem and come to different 
conclusions, e.g., for the Casimir-Polder force acting between an atom and 
a flat plasma sheet.
\\
This work was supported by the research funding from the EC's Sixth Framework Programme within the STRP project "PARNASS" (NMP4-CT-2005-01707).\footnote{Any and all views expressed or implied in this paper are exclusively those of authors and the European Community is not liable for any use that may be made of the information contained therein.}

\section*{Appendix}
In this appendix we consider a spherical plasma shell in order to give another illustration of the statement made at the end of section 4 that the excitations of the electromagnetic field and of the plasma have the same spectrum. We start from Eq.\Ref{sK0} which is the kernel of the action of the displacement field after integrating out the electromagnetic field and which be means of Eq.\Ref{eomxi} defines the equation of motion of the displacement field. First of all we have to consider the current $j_\mu(x)$, \Ref{j2}. The gradient entering $j_0$ contains derivatives in the tangential directions only, hence with $\nabla_i=n_i \pa_r-\frac{i}{r}(n\times L)_i$ where $n$ is the normal to the sphere and $L$ is the orbital momentum operator, we get $j_0=e\delta(r-R)\frac{-i}{R}(n\times L)_i \xi_i$. Further it is meaningful to introduce polarizations for the displacement field,
\be\label{Pi}P_i^{(1)}=L_i\frac{1}{\sqrt{L^2}}, \qquad P_i^{(2)}=(n\times L)_i\frac{1}{\sqrt{L^2}},
\ee
which form a basis. The corresponding expansion of the displacement fields reads
\be\label{}\xi_i(z)=\sum_{s=1}^2\ P_i^{(s)} \ \xi^{(s)}(z) \ .
\ee
We will see that the polarization $s=1$ couples to the TE-mode of the electromagnetic field and $s=2$ to the TM-mode. Now we consider the coupling with the electromagnetic field given in general form by  Eq.\Ref{H1}. In the spherical case we have instead of \Ref{H2} now 
\be\label{Hsp}H_{\mu i}=e\delta(r-R)\left(\delta_{\mu 0}\frac{-i}{R}(n\times L)_i-\delta_{\mu i}\pa_0\right).
\ee
Being  inserted \ into \Ref{H1}, the $H_{\mu i}$ appear in the combinations
\bea\label{}
H_{\mu i} P_i^{(1)}&=&-e \pa_0 \ E_\mu ^{(1)}
    \nn \\
H_{\mu i} P_i^{(2)}&=&-e \pa_0 \ E_\mu ^{(2)} \ ,
\eea
where we introduced the notations
\be\label{}
E_\mu ^{(1)}=\left(\begin{array}{c} 0 \\ \vec{L}\end{array}\right)_{\! \mu}
            \frac{1}{\sqrt{L^2}} \ ,
\qquad
E_\mu ^{(2)}=\left(\begin{array}{c} \frac{1}{R}L^2 \\
    i\pa_0 (\vec{n}\times\vec{L})\end{array}\right)_{\! \mu}\frac{1}{\Gamma \sqrt{L^2}} \ 
\ee
with $\Gamma=\sqrt{-\pa_0^2+L^2/R^2}$. These vectors $E_\mu^{(s)}$ 
are orthogonal and normalized like the corresponding ones in \Ref{Ems} in the flat case.
We note that these vectors up to differences in the notations coincide with those introduced in \cite{Bordag:1998sw}.

The kernel $^S\!K^0$, defined in Eq.\Ref{sK0}, carries indices like the displacement vectors $\xi_i$. We project it onto the polarizations \Ref{Pi} and  define
\be\label{sK0sp1}
\left(^S\!K^0\right)^{(s)}={P^\dagger}_i^{(s)} \ \left({^S\!K^0}\right)_{ij} \ P_j^{(s)} \ .
\ee
Here we took into account diagonalness which holds for obvious symmetry reasons.
Inserting into \Ref{sK0} we get 
\be\label{sK0sp2}
\left(^S\!K^0\right)^{(s)}= 
K^0-\left(H_{\mu i}P_i^{(s)}\right)^\dagger \ D_{\mu\nu}(z,z') \left(H_{\nu j}P_j^{(s)}\right) \ .
\ee
Here we took into account that $K^0=-m\pa_0^2$, Eq.\Ref{K0}, remains unchanged and that $K^{-1}$ is the free space photon propagator \Ref{Dmn} which we represent now in the form
\be\label{Dsp}
 D_{\mu\nu}(z,z') = g_{\mu\nu}\int\frac{dk_0}{2\pi}\ e^{ik_0(z_0-z'_0)}\sum_{lm}\ Y_{lm}(\Om) \ d_l(r,r') \ Y_{lm}^*(\Om')
\ee
with
\be\label{dl}d_l(r,r')=ik_0j_l(k_0 r_<) h^{(1)}_l(k_0 r_>)\ .
\ee
We note that we use Lorentz gauge ($\al=1$) but any other choice would result in the same expressions for $\left(^S\!K^0\right)^{(s)}$. Traced back to \Ref{j2} this is due to current conservation. 

Now we need to simplify \Ref{sK0sp2}. We define the corresponding quantities in momentum representation,
\be\label{}
\left(^S\!K^0\right)^{(s)}=
\int\frac{dk_0}{2\pi}\ e^{ik_0(z_0-z'_0)}\sum_{lm}\ Y_{lm}(\Om) \ \left(^S\!K^0\right)^{(s)}_{lm} \ Y_{lm}^*(\Om')
\ee
(note that $\left(^S\!K^0\right)^{(s)}$ is defined on the sphere).
For the polarization $s=1$ this is simple because the momentum operator $L$ commutes with the propagator and we get simply 
\bea\label{}
\left(^S\!K^0\right)^{(1)}_{lm} &=& mk_0^2  +  e^2 R^2 k_0^2 d_l(R,R)
\nn \\ &\equiv &mk_0^2 \ g_l^{(1)}(k_0)
\eea
with
\be\label{} g_l^{(1)}(k_0)=\left(1+\Om R^2 d_l(R,R)\right). 
\ee
The corresponding calculation for  the polarization $s=2$ is a bit more involved because the operator $(n\times L)$ does not commute with the propagator. In the basis of the orbital momentum eigenfunctions we have to consider
\be\label{}<lm|{E^\dagger}^{(2)}_\mu \ g_{\mu\nu} D(x-x') E_\nu^{(2)}| lm >.
\ee
The corresponding calculation was carried out in \cite{Bordag:1998sw} using known formulas from quantum mechanics and the result is
\bea\label{}
\left(^S\!K^0\right)^{(2)}_{lm} &=& mk_0^2  +  e^2 \frac{R^2}{L^2}\left(\frac{L^4}{r^2}d_l(R,R)
-<lm|{(n\times L)^\dagger_i} \   D(x-x') (n\times L)_i| lm >\right)
\nn \\ &\equiv &mk_0^2 \ g_l^{(2)}(k_0)
\eea
with
\be\label{}
g_l^{(2)}(k_0)=
1-\frac{\Om}{k_0^2}\left(\frac{L^2}{R^2}d_l(R,R)
        -k_0^2\frac{(l+1)d_l(R,R)+ld_l(R,R)}{2l+1}\right) \ .
\ee
Using the recursion relations for the Bessel functions the last line can be rewritten in terms of derivatives of the Riccati-Bessel functions,
\be\label{}
g_l^{(2)}(k_0)= 1+\frac{i\Om}{k_0} \ \hat{j}_l'(k_0 R) \hat{h}_l'(k_0 R) \ .
\ee
We finish this part with the remark that the functions $g_l^{(s)}(k_0)$ are the same as the Jost functions for the corresponding scattering problem  for the electromagnetic field in a spherical plasma shell. The corresponding scattering phase shifts are given in \cite{BV}. 

Returning to Eq.\Ref{sK0} it is clear that after expanding the displacement fields in the orbital momentum basis the equations of motion become algebraic ones like in the case of flat plasma sheet and that the spectrum is defined by the zeros of the functions $g_l^{(s)}(k_0)$. In this way, for a spherical shell, we demonstrated that the excitation spectrum of the electromagnetic field and of the displacement field are the same.
Further, since it is known that these functions do not have zeros for real $k_0$  we come to the conclusion that the displacement field does not have plasmon excitations. We conclude this appendix with the remark that the relation between the spectra of the excitations of the electromagnetic field and the displacement field is, of course, more involved. What we showed here is merely that the excitations of the displacement field are also excitations of the electromagnetic field since the zeros of $^S\!k^0$ \Ref{sK0} are poles of $^S\!D$ \Ref{sD}. But the electromagnetic field   has more solutions, namely the scattering states, which do not have a correspondence in the spectrum of the displacement field.

We remark, that of course, since we were looking for solutions with real $\om$, we have in mind excitations not decaying in time. Further we would like to mention that the absence of excitations found here is not in contradiction to the excitations found in \cite{Barton:1991pb} since those were obtained in an approximation which was nonrelativistic from the outset.

\bibliographystyle{unsrt}\bibliography{../../../Literatur/Bordag,../../../Literatur/articoli,../../../Literatur/libri}

\begin{thebibliography}{10}

\bibitem{Bordag:2004dn}
M.~Bordag.
\newblock Reconsidering the quantization of electrodynamics with boundary
  conditions and some measurable consequences.
\newblock {\em Phys. Rev.}, D70:085010, 2004.

\bibitem{Fetter73}
A.~L. Fetter.
\newblock {Electrodynamics of a Layered Electron-Gas.1. Single Layer}.
\newblock {\em Annals of Physics}, 81(2):367--393, 1973.

\bibitem{Barton:1991pb}
Gabriel Barton and Claudia Eberlein.
\newblock {Plasma spectroscopy proposed for C$_{60}$ and C$_{70}$}.
\newblock {\em J. Chem. Phys.}, 95:1512--1517, 1991.

\bibitem{Maksimenko2002}
S.A. Maksimenko and G.Ya. Slepyan.
\newblock {Electrodynamics of Carbon Nanotubes}.
\newblock {\em J. Comm. Tech. Electron.}, 47:235--252, 2002.

\bibitem{BIV}
G.~Barton.
\newblock Casimir's spheres near the coulomb limit: energy density, pressures
  and radiative effects.
\newblock {\em Journal of Physics A-Mathematical and General},
  37(11):3725--3741, 2004.

\bibitem{Bordag:2005by}
M.~Bordag.
\newblock {The Casimir effect for thin plasma sheets and the role of the
  surface plasmons}.
\newblock {\em J. Phys. A: Math. Gen.}, 39:6173--6185, 2006.

\bibitem{BORDAG2006D}
M~Bordag, B.~Geyer, G.~L. Klimchitskaya, and V.~M. Mostepanenko.
\newblock {Lifshitz-type formulas for graphene and single-wall carbon
  nanotubes: van der Waals and Casimir interactions}.
\newblock {\em Phys.Rev.B}, 74:205431, 2006.

\bibitem{Bordag:1985zk}
M.~Bordag, D.~Robaschik, and E.~Wieczorek.
\newblock {Quantum Field Theoretic Treatment of the Casimir Effect}.
\newblock {\em Ann. Phys.}, 165:192, 1985.

\bibitem{Hennig:1990ap}
D.~Hennig and D.~Robaschik.
\newblock Feynman propagators in the presence of external potentials
  concentrated on planes.
\newblock {\em Phys. Lett.}, A151:209--214, 1990.

\bibitem{Bordag:1992cm}
M.~Bordag, D.~Hennig, and D.~Robaschik.
\newblock {Vacuum energy in quantum field theory with external potentials
  concentrated on planes}.
\newblock {\em J. Phys. A}, A25:4483, 1992.

\bibitem{WIRZBA2006}
Aurel Bulgac, Piotr Magierski, and Andreas Wirzba.
\newblock {Scalar Casimir effect between Dirichlet spheres or a plate and a
  sphere}.
\newblock {\em Phys. Rev.}, D73:025007, 2006.

\bibitem{Bordag:1984ht}
M.~Bordag.
\newblock {On the Canonical Quantization of QED with Boundary Conditions}.
\newblock 1984.
\newblock JINR-P2-84-115.

\bibitem{Bordag:1998sw}
M~Bordag and J~Lindig.
\newblock {Radiative correction to the Casimir force on a sphere}.
\newblock {\em Phys. Rev.}, D58:045003, 1998.

\bibitem{BV}
G.~Barton.
\newblock {Casimir effects for a flat plasma sheet: I. Energies}.
\newblock {\em J. Phys.}, A38(13):2997--3019, 2005.

\bibitem{Cp48}
H.~B.~G. Casimir and D.~Polder.
\newblock {The Influence of Retardation on the London-Van der Waals Forces}.
\newblock {\em Physical Review}, 73(4):360--372, 1948.

\bibitem{Barton:1970hw}
G.~Barton.
\newblock Quantum electrodynamics of spinless particles between conducting
  plates.
\newblock {\em Proc. Roy. Soc. Lond.}, A320:251--275, 1970.

\bibitem{Hinds}
E.A. Hinds.
\newblock {Cavity Quantum Electrodynamics}.
\newblock {\em {in: Advances in Atomic, Molecular and Optical Physics}},
  28:237--89, 1991.
\newblock Editor: D.Bates, Academic Press, Boston, 1991.

\bibitem{BVI}
G.~Barton.
\newblock {Casimir effects for a flat plasma sheet: II. Fields and stresses}.
\newblock {\em Journal of Physics A-Mathematical and General},
  38(13):3021--3044, 2005.

\end{thebibliography}
\end{document}